\documentclass[a4paper,twocolumn,11pt,accepted=2023-02-13]{quantumarticle}
\pdfoutput=1
\usepackage{graphicx,comment}
\usepackage{amsfonts}
\usepackage{amsmath,amssymb}
\usepackage{nth}
\usepackage{bm}
\usepackage{color}
\usepackage{epstopdf}
\usepackage{epsfig}
\usepackage{tikz}
\usepackage{pgfplots}
\usepackage{qcircuit}
\usepackage{float}
\usepackage[numbers,sort&compress]{natbib}
\pgfplotsset{
table/search path={csv/},
}
\usetikzlibrary{pgfplots.groupplots}
\usepackage[font=small,
   justification=justified,
   format=plain]{caption} 
\usepackage{subfig}
\usepackage{verbatim}
\usepackage{hyperref}
\usepackage{multirow}
\usepackage{tabularx}
\usepackage{wrapfig}
\hypersetup{
     colorlinks   = true,
    citecolor    = blue
}
\usepackage{lineno}
\usepackage[thicklines]{cancel}
\usepackage{color, colortbl}
\definecolor{LightCyan}{rgb}{0.88,1,1}
\usepackage{url}
    
{\rm }

\def\be{\begin{equation}}
 \def\ee{\end{equation}}
 \def\bea{\begin{eqnarray}}
 \def\eea{\end{eqnarray}}

\def\2{\frac{1}{2}}
\def\4{\frac{1}{4}}

\newcommand{\bra}[1]{\left\langle #1 \right|}

\newcommand{\ket}[1]{\left| #1 \right\rangle}

\definecolor{red}{rgb}{1,0.,0}

\newcommand{\Yb}{\textsuperscript{171}Yb\textsuperscript{+}}
\newcommand{\state}[3]{\textsuperscript{#1}#2\textsubscript{#3}}

\begin{document}


\title{Characterizing and mitigating coherent errors in a trapped ion quantum processor using hidden inverses}

\author{Swarnadeep Majumder}
\email{swarnadeep.majumder@duke.edu}
\thanks{Current affiliation: IBM T.J. Watson Research Center, Yorktown Heights, New York 10598, USA}
\affiliation{Duke Quantum Center, Duke University, Durham, NC 27701, USA}%
\affiliation{Department of Electrical and Computer Engineering, Duke University, Durham, NC 27708 USA}%

\author{Christopher G. Yale}
\email{cgyale@sandia.gov}
\affiliation{Sandia National Laboratories, Albuquerque, NM 87123, USA}

\author{Titus Morris}
\email{morristd@ornl.gov}
\affiliation{Quantum Information Science Section, Oak Ridge National Laboratory, Oak Ridge, TN 37831, USA}

\author{Daniel S. Lobser}
\affiliation{Sandia National Laboratories, Albuquerque, NM 87123, USA}

\author{Ashlyn D. Burch}
\affiliation{Sandia National Laboratories, Albuquerque, NM 87123, USA}

\author{Matthew N. H. Chow}
\affiliation{Sandia National Laboratories, Albuquerque, NM 87123, USA}
\affiliation{Department of Physics and Astronomy, University of New Mexico, Albuquerque, NM 87131, USA}
\affiliation{Center for Quantum Information and Control, University of New Mexico, Albuquerque, NM 87131, USA}

\author{Melissa C. Revelle}
\affiliation{Sandia National Laboratories, Albuquerque, NM 87123, USA}

\author{Susan M. Clark}
\affiliation{Sandia National Laboratories, Albuquerque, NM 87123, USA}

\author{Raphael C.\ Pooser}
\affiliation{Quantum Information Science Section, Oak Ridge National Laboratory,
  Oak Ridge, TN 37831, USA}

\begin{abstract}
Quantum computing testbeds exhibit high-fidelity quantum control over small collections of qubits, enabling performance of precise, repeatable operations followed by measurements. Currently, these noisy intermediate-scale devices can support a sufficient number of sequential operations prior to decoherence such that near term algorithms can be performed with proximate accuracy (like chemical accuracy for quantum chemistry problems). While the results of these algorithms are imperfect, these imperfections can help bootstrap quantum computer testbed development. Demonstrations of these algorithms over the past few years, coupled with the idea that imperfect algorithm performance can be caused by several dominant noise sources in the quantum processor, which can be measured and calibrated during algorithm execution or in post-processing, has led to the use of noise mitigation to improve typical computational results. Conversely, benchmark algorithms coupled with noise mitigation can help diagnose the nature of the noise, whether systematic or purely random. Here, we outline the use of coherent noise mitigation techniques as a characterization tool in trapped-ion testbeds. We perform model-fitting of the noisy data to determine the noise source based on realistic physics focused noise models and demonstrate that systematic noise amplification coupled with error mitigation schemes provides useful data for noise model deduction. Further, in order to connect lower level noise model details with application specific performance of near term algorithms, we experimentally construct the loss landscape of a variational algorithm under various injected noise sources coupled with error mitigation techniques. This type of connection enables application-aware hardware codesign, in which the most important noise sources in specific applications, like quantum chemistry, become foci of improvement in subsequent hardware generations.

%
\end{abstract}

\maketitle

\section{Introduction}
Small algorithms on quantum computing testbed devices can help diagnose performance problems when the resulting data is viewed within the context of dominant low-level noise sources. In near-term pre-fault-tolerant devices, coherent noise (such as over- or under-rotation errors) can lead to an array of incorrect results at the algorithmic level, including incorrect expectation values in time evolution simulations or slow convergence and incorrect parameter determination in variational algorithms. Here, we use prototypical variational quantum chemistry calculations as a diagnostic algorithm to probe coherent noise on a trapped ion testbed platform. Two methods of noise mitigation - randomized compiling~\cite{wallman_noise_2016} and the application of hidden inverse gates~\cite{zhang_hidden_2022} - are used to both characterize and mitigate time-dependent coherent noise in test circuits. We present simulations of one and two-qubit noise models on trapped ion quantum devices, validated by experimental data, to verify this hypothesis.

Quantum errors in a quantum computer can be thought of as any evolution of the qubits that differs from the ideal intended operation. There are different ways one may categorize quantum errors, but for the purpose of this work we limit our discussion to two distinct types of errors: incoherent error and coherent error. Incoherent errors map pure states to mixed states resulting in loss of purity often caused by classical noise. Some well known quantum channels such as depolarization and dephasing are examples of incoherent errors.  Conversely, coherent error can be represented as a single unitary operator and maps pure states to pure states. Such an error usually comes from miscalibration or drift out of calibration of the control system used to drive the qubit operations. If one has a good understanding of the system interaction Hamiltonian, it is often possible to describe coherent errors as additive or multiplicative terms in the control parameters. This allows one to use targeted characterization procedures to learn about the error parameters which in turn can be used for better calibration. While fault tolerant protocols using a quantum error correction (QEC) code have been demonstrated recently~\cite{egan_fault-tolerant_2021,krinner_realizing_2021,ryan-anderson_realization_2021}, noisy intermediate scale quantum (NISQ) devices are unable to take advantage of QEC because of high noise levels and small system sizes. For NISQ devices, one can try to mitigate these errors to minimize their impact on the quantum computer's output; such protocols are collectively knows as EM (error mitigation) protocols.
Multiple quantum characterization, verification, and validation (QCVV) protocols exist for characterizing the noise channels and error models discussed here, including gate set tomography~\cite{blume-kohout_robust_2013}, randomized benchmarking~\cite{johnson_demonstration_2015}, generalized model fitting~\cite{nielsen_efficient_2021}, and many others. While these frameworks are very general, typically allowing for deduction of a diverse array of noise models, a large quantity of experimental data can be required to learn the models in some cases. The exclusive access to many quantum computing platforms required to obtain the data is not always readily available. Recent applications of various EM techniques in the NISQ era provide a promising middle ground approach~\cite{nation_scalable_2021,kim_scalable_2021,peters_perturbative_2021,strikis_learning-based_2021,piveteau_error_2021,larose_mitiq_2020,zhang_error-mitigated_2020,czarnik_error_2020, Suzuki2022}.  Zero noise extrapolation (ZNE) and probabilistic error cancellation (PEC) \cite{Temme2017, berg2022probabilistic}  are two other error mitigation protocols that have been used in various applications run on superconducting qubit devices. ZNE and PEC are more well suited for incoherent errors but coherent noise can be twirled into Pauli noise and these protocols can be effectively used as demonstrated in \cite{berg2022probabilistic} for PEC.  Essentially, the efficacy of an EM routine is an indicator of the presence of a specific type of noise present in the machine, and with sufficient \textit{a priori} information these techniques can be used to characterize noise in a quantum processor. 



\section{Hidden inverses protocol to mitigate coherent errors}
Hidden inverses (HI) \cite{zhang_hidden_2022, HI_vicente, Kubra2021AQT} is an EM protocol that addresses coherent errors via noise cancellation. HI does not require any additional gates or any post processing. The key idea behind HI is that a few of the common unitaries used in quantum computing are self-adjoints. These self-adjoint unitary operators are constructed from primitive gates which themselves are not self-adjoints. As an example if we want to apply a certain gate G = ABC, we can either apply G = ABC or $G^\dagger=$ $C^\dagger B^\dagger A^\dagger$. While in the absence of noise, G and $G^\dagger$ implement the same physical operation; this is not true in general when A, B, C, $A^\dagger$, $B^\dagger$ and $C^\dagger$ are subjected to errors. Given the freedom to apply G or $G^\dagger$, a compiler needs to choose which one to apply based on other gates in the vicinity within a quantum circuit along with an understating of the noise process. Developing compiler logic for picking the optimal configuration given an arbitrary circuit and a noise model is an open problem and one that we will not discuss in this paper. Instead, we focus our attention on demonstrating hidden inverses's performance with hand-crafted circuits. It has been shown theoretically that hidden inverses can improve circuit performance for arbitrary circuit sizes as long as the right configuration is chosen based on the noise model. Choosing the right configuration is a local optimization problem as such the complexity is independent of the circuit size. We refer interested readers to \cite{zhang_hidden_2022} for more theoretical and simulation results of the hidden inverses protocol.   

\begin{figure}
\resizebox{.9\columnwidth}{!}{
\centerline{
\raisebox{-1.4em}{\hspace{1mm}$\mathrm{CNOT}$\hspace{1mm}}
\Qcircuit @C=0.5em @R=1.35em @!R{
            &\qw&\ctrl{1}&\qw\\
            &\qw&\targ&\qw
            }
\raisebox{-1.4em}{\hspace{1mm}$\equiv$\hspace{1mm}}
\Qcircuit @C=0.5em @R=0.5em @!R {
 	& \gate{Y(\frac{\pi}{2})}	& \multigate{1}{XX(\frac{\pi}{2})} 	& \gate{X(-\frac{\pi}{2})} 	& \gate{Y(-\frac{\pi}{2})}	& \qw\\
& \qw & \ghost{XX(\frac{\pi}{2})} & \gate{X(-\frac{\pi}{2})}	& \qw&\qw
}}
}
\resizebox{.9\columnwidth}{!}{
\centerline{
\raisebox{-1.4em}{\hspace{1mm}$\mathrm{CNOT}^\dagger$\hspace{0.0mm}}
\Qcircuit @C=0.5em @R=1.35em @!R{
            &\qw&\ctrl{1}&\qw\\
            &\qw&\targ&\qw
            }
\raisebox{-1.4em}{\hspace{1mm}$\equiv$\hspace{1mm}}
\Qcircuit @C=0.5em @R=0.5em @!R {
& \gate{Y(\frac{\pi}{2})}&\gate{X(\frac{\pi}{2})}&\multigate{1}{XX(-\frac{\pi}{2})}& \gate{Y(-\frac{\pi}{2})}&\qw\\
&\qw&\gate{X(\frac{\pi}{2})}& \ghost{XX(-\frac{\pi}{2})}&\qw&\qw
}}
}
	\caption[]{Standard and Hermitian conjugated decompositions of $\mathrm{CNOT}$ gate with native trapped ion quantum operations consisting of single qubit gates
	and M\o lmer-S\o rensen interactions. }
		\label{fig:cnot_decomp}
\end{figure}
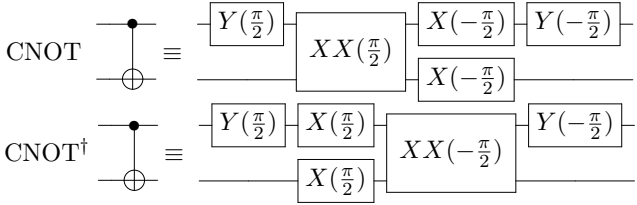

Hadamard (H) and controlled-NOT (CNOT) are two self-inverse unitary operations widely used in quantum computing. We can decompose their standard or native configurations into trapped-ion native single qubit gates and M\o lmer-S\o rensen (MS) interactions (Fig \ref{fig:H_decomp}, \ref{fig:cnot_decomp}).

\begin{figure}
\centerline{
\Qcircuit @C=1em @R=.7em @!R{
            &\gate{H}&\qw
            }
\raisebox{0em}{\hspace{1mm}$\equiv$\hspace{1mm}}
\Qcircuit @C=1em @R=.7em {
   & \gate{Y(-\frac{\pi}{2})} & \gate{X(\pi)} & \qw 
}}

\centerline{
\Qcircuit @C=1em @R=.7em @!R{
            &\gate{H^\dagger}&\qw
            }
\raisebox{0em}{\hspace{1mm}$\equiv$\hspace{1mm}}
\Qcircuit @C=1em @R=.7em {
   & \gate{X(-\pi)} & \gate{Y(\frac{\pi}{2})} & \qw 
}}
\caption[]{Standard and Hermitian conjugated decompositions of $\mathrm{H}$ gate with native trapped ion quantum operations consisting of single qubit gates}
		\label{fig:H_decomp}
\end{figure}
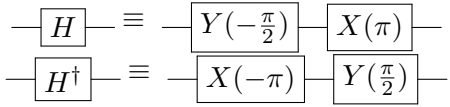

\section{Experimental implementation of quantum circuits on a trapped ion processor}
\label{sec:qscout_exp}
We investigate these hidden inverse protocols on the Quantum Scientific Computing Open User Testbed (QSCOUT), a room-temperature quantum processor based on trapped ions \cite{clark-2021} housed at Sandia National Laboratories. This investigation consisted of either one or two qubits, in which the qubit states comprise the hyperfine `clock' transition of a \Yb{} ion, \state{2}{S}{1/2}~$|$F=0, $m_F =0\rangle$ ($\ket{0}$) and $|$F=1, $m_F=0\rangle$ ($\ket{1}$) \cite{olmschenk-2007}. The ions are trapped in a radio-frequency (RF) pseudopotential generated on a microfabricated surface electrode trap, the Sandia High Optical Access (HOA2.1) trap \cite{maunz-2016}, with radial frequencies, $\omega_{r,i}/2\pi =$ 2.2 and 2.5~MHz, and an axial frequency, $\omega_{a}/2\pi =$ 700~kHz. This axial frequency sets the spacing of the ions, 4.5~$\mu$m, to match the spacing of the individual addressing beams. Each ion is then imaged in a separate core of a multicore fiber array via magnification optics to match the fiber core spacing to the 4.5~$\mu$m separation. The ions are cooled and detected along their \state{2}{S}{1/2} to \state{2}{P}{1/2} transition using 370~nm light.

All gates performed in the system are based on Raman transitions from a pulsed 355~nm laser \cite{hayes-2010} with beams in a counter-propagating configuration, consisting of a wide `global' beam illuminating all ions and individual-ion addressing beams generated via a multi-channel acousto-optic modulator (AOM) \cite{debnath-2016} counter to the global beam. Both the multi-channel AOM and the single-channel AOM for the global beam are driven by a custom coherent control system, Octet, providing two tones for each AOM channel. 

Gates are fully parameterized within the QSCOUT system, providing both arbitrary phase and rotation angle for single- and two-qubit gates.  Single-qubit gates are driven using two Raman tones on the appropriate individual beam in a co-propagating Raman configuration. Our two-qubit MS gate is generated with two tones on each participating ion's individual beam and another tone on the global beam.  The three tones are required to form the appropriate Raman transitions symmetrically detuned from a red and blue motional sideband. The MS gate as implemented in QSCOUT additionally has a series of single-qubit basis transformations to eliminate phase instabilities between the counter- vs. co- propagating configurations. The bare MS($\theta$) gate is an XX interaction, XX$(\theta) = e^{-i\frac{\theta}{2} \sigma_{X} \otimes \sigma_{X}}$, that exists in a counter-propagating basis. By surrounding the two-qubit gate with the appropriate counter-propagating single-qubit gates, we transform the XX interaction into a ZZ interaction to eliminate phase instabilities with the co-propagating single-qubit gates \cite{lee-2005}. The gate is then further surrounded with co-propagating single-qubit gates to complete the transformation back into an XX interaction within the co-propagating basis.

The physical single-qubit gates' ($R_X$ and $R_Y$) amplitudes are square-pulsed and gapless, meaning there is no ``off'' period between single-qubit gates.  Their rotation angle is determined by the duration of the pulse. $R_Z$ gates are virtual, treated as cumulative phase shifts by Octet. 

The MS gate amplitude is Gaussian-shaped, and the angle of rotation is controlled by the global beam amplitude while accounting for distortions and saturation in the amplifier and AOM. A MS gate with a negative rotation angle is achieved by flipping the phase on one of the two qubits by $\pi$ radians.  For all gates, phases are set by the relative phase difference between two of the RF tones applied to the ion. The MS gate also incurs an AC Stark shift which is cancelled through a dynamically evolving virtual Z rotation applied throughout the duration of the MS gate pulse.
This ability to fully control both the phase and rotation angle for single- and two-qubit gates is needed for these investigations to intentionally introduce static coherent rotation and phase errors. Typical physical single-qubit gate fidelities for a rotation angle of $\pi/2$ are estimated to be 99.5$\pm$0.3$\%$, while two-qubit XX($\pi/2$) are 97$\pm$1$\%$. 

Additionally, to cool the ions to near the motional ground state, the ions are both Doppler cooled and resolved-sideband cooled \cite{deslauriers-2004}. Typically, 60 loops of cooling are performed on all relevant sidebands to reach the minimum motional state; however, the number of cooling loops can be reduced so the ion will begin the desired circuit in a higher motional mode, inducing additional errors. These errors include an overall static coherent under-rotation error and shot-to-shot coherent errors. At very small numbers of cooling loops ($\lesssim$ 10 loops), there is an increase in the population of non-desirable qubit states - i.e. for a state beginning in $\ket{00}$, an MS gate ideally creates a state somewhere within the $\ket{00}$ and $\ket{11}$ basis, but instead there are nonzero populations in the $\ket{10}$ and $\ket{01}$ states. These unwanted populations can be as large as $\sim 5\%$ when no cooling loops are applied. 

The gate sequences are all programmed using Just Another Quantum Assembly Language (Jaqal), a flexible programming language designed to support the underlying hardware of the QSCOUT system \cite{morrison-2020}.  Since Jaqal contains the fully parameterized gateset, we can introduce intentional coherent errors directly at the Jaqal level as different rotation angles or phases, rather than underlying waveforms, described by JaqalPaw (Pulses and Waveforms), the pulse-level counterpart to Jaqal \cite{JaqalPaw}. On the other hand, errors introduced by insufficient cooling of the ion are instituted by adapting the cooling cycle in the waveform language JaqalPaw. 

\begin{figure}[ht]
    \centering
    \includegraphics[width=\columnwidth]{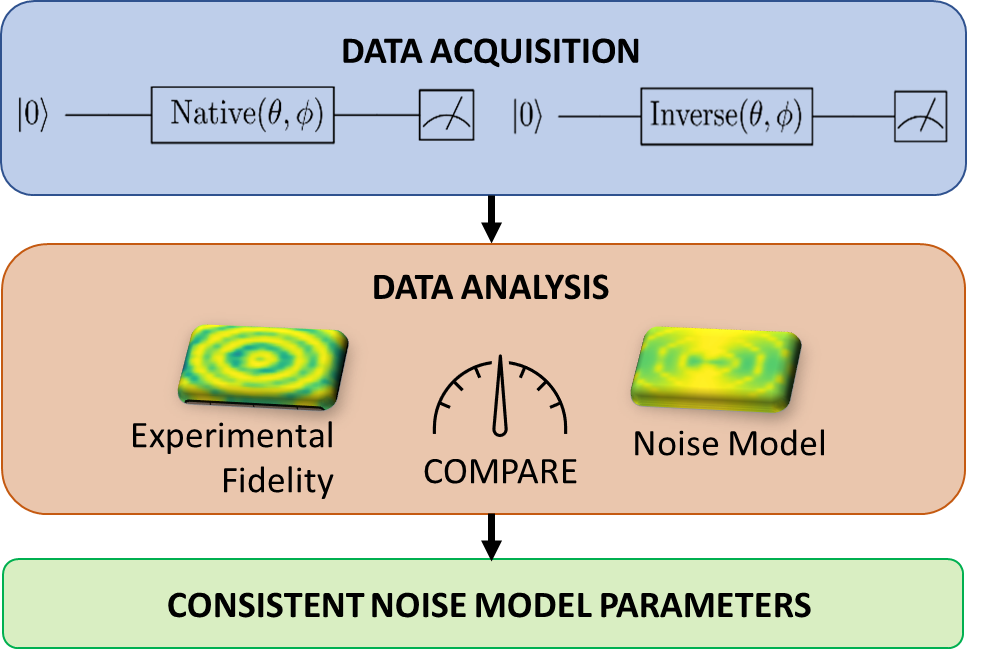}
    \caption{Illustration of noise characterization using single qubit hidden inverses.}  
    \label{HXZH}
\end{figure}

\section{Demonstration of hidden inverses as an error characterization experiment with single qubit gates}

\subsection{Experimental design}
We use H, H$^\dagger$, and a few single qubit rotations to develop an error characterization experiment. Here we summarize the key insight of the protocol. If we compare outputs of two circuits (1) $H-H$ (native circuit) and (2) $H- H^\dagger$ (inverted circuit) under coherent noise model that inverts with the inverse of the gate, we find circuit (2) will result in error cancellation while circuit (1) will amplify the error. In order to investigate different sources of coherent errors in the system, we inject parameterized single qubit rotations in between the gates as follows (1) $H- X(\Theta)-Z(\Phi)- H$, $H -Y(\Theta)- Z(\Phi)- H$  or (2) $H- X(\Theta) -Z(\Phi)- H^\dagger$, $H -Y(\Theta)- Z(\Phi)- H^\dagger$. When we sweep over $\{\Theta, \Phi\}$, we will amplify or cancel various types of coherent errors in the native and inverted circuits. For some choices of $\{\Theta, \Phi\}$ the inverse circuits will outperform the native circuits, while for others the native circuits will be a better choice. Aided with numerical simulations, patterns in the fidelity (or population) plots of the various circuits in phase space of $\{\Theta, \Phi\}$ will reveal the underlying noise process.

There are two stages in this experiment (i) data collection stage and (ii) data analysis stage. 
In the data collection stage, the circuits were run 100 times before a measurement, 
i.e. (1) [$H -X(\Theta)- Z(\Phi)- H$, $H- Y(\Theta) -Z(\Phi)- H$]$^{100}$  or (2) [$H- X(\Theta)- Z(\Phi)- H^\dagger$, $H- Y(\Theta)- Z(\Phi)- H^\dagger$]$^{100}$ to amplify any effects of cancellation. 21 equally spaced points were chosen in between $-\frac{\pi}{36}\leq (\Theta, \Phi) \leq \frac{\pi}{36} $ for a total of 441 points in a 2D-grid for each circuit. Each data set was taken with random ordering in $\Theta$ and $\phi$ to average out effect of any long term drift. Outcomes were averaged over 200 shots taken back-to-back for a given $(\Theta, \Phi)$. 200 shots were chosen for sufficient averaging without significant impacts from system drift. This data was taken without dynamical decoupling or composite pulses. 

In order to characterize system parameter drift and verify the model, [$H- X(\Theta)- Z(\Phi)- H^\dagger$]$^{100}$ was also run multiple times back to back for four different cases (1) natural drift: no re-calibration was performed between experiments (2) re-calibration: re-calibration was performed between experiments (3) intentional amplitude noise injection (4) intentional phase noise injection.   

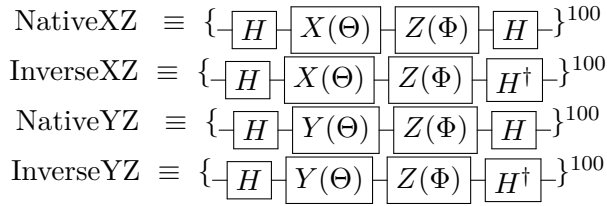
\begin{figure}
\centerline{
\raisebox{0em}{\hspace{1mm}$\mathrm{Native XZ}$\hspace{1mm}}

\raisebox{0em}{\hspace{1mm}$\equiv$\hspace{1mm}}
\{\Qcircuit @C=0.5em @R=0.5em @!R {
 	& \gate{H}	& \gate{X(\Theta)} 	& \gate{Z(\Phi)} 	& \gate{H}	& \qw 
}\}$^{100}$}

\centerline{
\raisebox{0em}{\hspace{1mm}$\mathrm{Inverse XZ}$\hspace{0.0mm}}

\raisebox{0em}{\hspace{1mm}$\equiv$\hspace{1mm}}
\{\Qcircuit @C=0.5em @R=0.5em @!R {
 	& \gate{H}	& \gate{X(\Theta)} 	& \gate{Z(\Phi)} 	& \gate{H^\dagger}	& \qw
}\}$^{100}$}

\centerline{
\raisebox{0em}{\hspace{1mm}$\mathrm{Native YZ}$\hspace{1mm}}

\raisebox{0em}{\hspace{1mm}$\equiv$\hspace{1mm}}
\{\Qcircuit @C=0.5em @R=0.5em @!R {
 	& \gate{H}	& \gate{Y(\Theta)} 	& \gate{Z(\Phi)} 	& \gate{H}	& \qw 
}\}$^{100}$}

\centerline{
\raisebox{0em}{\hspace{1mm}$\mathrm{Inverse YZ}$\hspace{0.0mm}}

\raisebox{0em}{\hspace{1mm}$\equiv$\hspace{1mm}}
\{\Qcircuit @C=0.5em @R=0.5em @!R {
 	& \gate{H}	& \gate{Y(\Theta)} 	& \gate{Z(\Phi)} 	& \gate{H^\dagger}	& \qw
}\}$^{100}$}

	\caption[]{Standard and Hermitian conjugated circuits for single qubit error characterization. \{\}$^{100}$ denotes the circuit is repeated 100 times before measurement. }
		\label{fig:hadamard_circuits}
\end{figure}

\subsection{Classical simulation and noise model}

 Outcomes from stage (1) were first converted into fidelity. To calculate fidelity, we performed noisy and ideal simulations of the quantum circuits. For the noisy simulation, we assume a physics-focused, single-qubit, ion-trap noise model parameterized by over-rotation error ($\epsilon$), phase error ($\phi$), and detuning error ($\delta$). In this model, ideal $X_{ideal}(\theta=\Omega t) = e^{-i\frac{\Omega t}{2} X}$ and $Y_{ideal}(\theta=\Omega t) = e^{-i\frac{\Omega t}{2} Y}$ where $\Omega$ is the Rabi frequency, become:
\begin{align}
X_{noisy} = e^{-i \frac{\Omega (1+\epsilon
) t}{2}(Cos(\phi) X + Sin(\phi) Y) + \frac{\delta t}{2} Z}\\
Y_{noisy} = e^{-i \frac{\Omega (1+\epsilon
) t}{2}(Cos(\phi) Y + Sin(\phi) X) + \frac{\delta t}{2} Z} \,.
\end{align}


\subsection{Result}
\begin{figure}
    \centering
    \includegraphics[width=1.0\columnwidth]{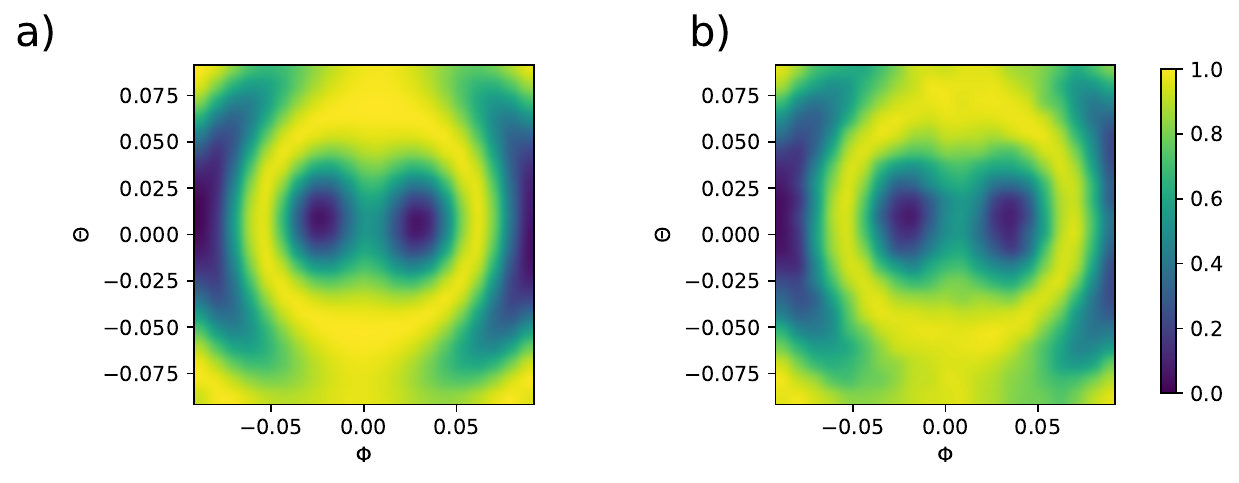}
    \caption{Phase space population (probability of measuring $\ket{0}$) after  applying [$H- X(\Theta)- Z(\Phi)- H^\dagger$]$^{100}$ on $\ket{0}$ as a function of \{$\Theta, \Phi\}$ for (a) simulation and (b) experiment}
    \label{fig:XZ-Hinv}    
\end{figure}

\begin{figure*}[t!]
    \centering
    \includegraphics[width=0.78\paperwidth]{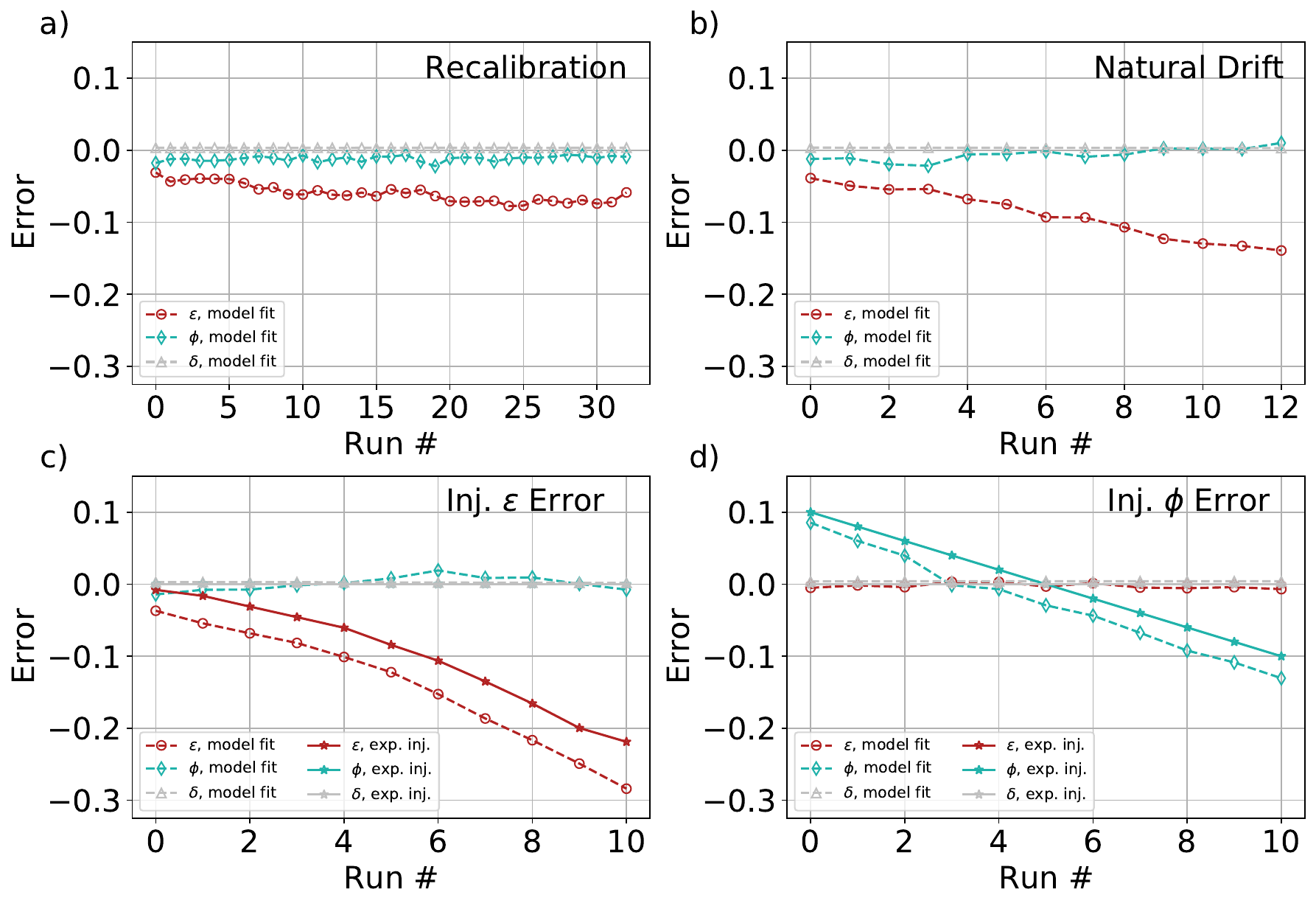}
    \caption{Error parameter drift} Error parameters as a function of number of runs are plotted. (a) is for the case where system parameters were calibrated in between runs and we find the estimated error parameters stable. (b) is for natural drift where re-calibration steps were skipped. We find signification drift in the estimated over-rotation (red) parameter. We then inject artificial noise in the form of amplitude error (c) and phase error (d). We find good agreement between the trends of the injected noise (solid line) and estimated parameters (dotted line). Offsets between the actual values of the injected noise and the estimated noise parameters are likely due to co-variances amongst the fitting parameters, $\epsilon$, $\phi$, and $\delta$.
    \label{Drift}
\end{figure*}

We found the best noise parameters by curve fitting the experimental data with noisy simulation using SciPy's non-linear least-square fitting algorithm \cite{SciPy-NMeth}. We used $\{\Theta, \Phi\}$ as the two-dimensional independent variables and the population as the dependent variable. Fig \ref{fig:XZ-Hinv}  displays the population phase space for [$H- X(\Theta)- Z(\Phi)- H^\dagger$]$^{100}$ circuit. In order to characterize drifting of error parameters in time, we plot the estimated error parameters as a function of number of runs in Fig \ref{Drift}. We find the variance of the parameter estimate from the curve-fit to be in the order of $10^{-4}$, hence they are omitted in the plots. Long term drifts in $\epsilon$ while $\phi$ and $\delta$ approximately stays constant. Interleaved re-calibration routine largely stabilizes $\epsilon$. With the addition of intentional amplitude and phase noise injection over many runs, we find the protocol is able to accurately track the error parameters.  


\section{Hidden inverses as an error mitigation protocol for applications: variational quantum eigensolvers  }
\subsection{Hamiltonian and Ansatz}
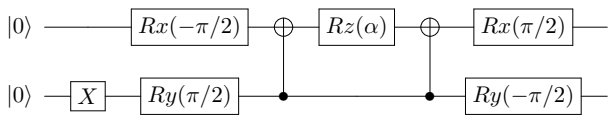
\begin{figure}[H]
    \centering
\resizebox{\columnwidth}{!}{
\Qcircuit @C=1.14em @R=1.44em {
  \push{\rule{1em}{0em}\rule{0em}{1em}} &\lstick{\ket{0}}  & \qw      & \gate{Rx(-\pi/2)} &\targ    &\gate{Rz(\alpha)} &\targ    &\gate{Rx(\pi/2)} &\qw\\
                                        &\lstick{\ket{0}}  & \gate{X} & \gate{Ry(\pi/2)}  &\ctrl{-1} & \qw             &\ctrl{-1} & \gate{Ry(-\pi/2)}&\qw}
}
    \caption{VQE tapered ansatz}
    \label{fig:qscout_vqe}
\end{figure}

Variational quantum eigensolvers (VQE) are a successful cornerstone of hybrid quantum-classical algorithms explored on NISQ era machines.  For a given system Hamiltonian $H$, and parameterized ansatz $\ket{\Psi(\alpha)}$, the ground state and its energy are obtained by classically minimizing the energy,
\begin{align}
E(\alpha) = \bra{\Psi(\theta)}H\ket{\Psi(\alpha)}\,.
\end{align}
Thus they form an ideal platform to explore the efficacy of mitigation algorithms like hidden inverse and randomized compiling.  In order to pursue this goal, we explore the VQE in the minimal basis for equilibrium molecular hydrogen. We use the Brayvi-Kitaev mapping, along with qubit tapering to yield a spin Hamiltonian of the following form, 

\small
\begin{align}
H = 0.304794*II+0.3555426*IZ-.485486*ZI \notag \\ 
+0.581232*ZZ+0.089500*(XX+YY)\,,
\end{align}
\normalsize
indicating that energy evaluation requires at most three circuit evaluations for ZZ, XX,and YY expectation values. We test the hidden inverse in this system in a prototypical one parameter two-qubit circuit ansatz in Fig \ref{fig:qscout_vqe}. 

This circuit is easily decomposed into native ion gates as in Fig \ref{fig:H_decomp}, where the default compilation uses the same construction for both CNOTs, and the HI implementation uses the CNOT$^\dagger$ construction for the second CNOT.  For the default and HI circuit, we swept over all angles $|\alpha| < \frac{\pi}{2}$, with 200 shots for each evaluation.
\subsection{Randomized Compiling}
We also wish to compare the benefit of HI method on the VQE ansatz (Fig. \ref{fig:qscout_vqe}) with other methods of coherent error mitigation, namely randomized compiling.  Randomized compiling is a powerful tool, where coherent errors of hard gates, typically those of the two-qubit gates, are suppressed by averaging over unitarily equivalent circuits of the two-qubit gates via application of easy ``twirling" gates.  In \cite{zhang_hidden_2022}, we simulated performance of randomized compiling and compared with hidden inverses for a set of noise models. We found hidden inverses can improve overall circuit fidelity in certain noise processes. In order to demonstrate how this fidelity improvement translates to improvement in algorithmic performance experimentally, we compare hidden inverses with randomized compiling for our VQE problem.  For application to our problem, we identified the CNOT as our hard gates, and the Pauli operators as our twirling group, yielding a total of 256 unique circuit evaluations.  We found that randomly choosing 10 circuits from this grouping was sufficient for converged results.  For each of the random circuits, they were repeated 20 times to yield the same number of shots to compare with the default and HI circuits. All three circuit variations (default, HI, and RC) were taken interleaved with one another on the experiment, and 200 shots per variation was chosen to limit system drift while maintaining sufficient averaging to isolate differences in results.

\subsection{Purification}
To further explore mitigation strategies in conjunction with the circuit level strategies already presented, we utilized fermionic density matrix purification first demonstrated in quantum computing in \cite{chem-bench}.  Here, instead of directly evaluating the energy based on measured Pauli expectations, we note that the 2-dimensional reduced fermionic density matrix, $\rho$ can also be established from the already measured expectation values.

\small
\begin{align}
   \rho &= \begin{pmatrix} \bra{01}\Psi \rangle \langle\Psi\ket{01} & \bra{01}\Psi \rangle \langle\Psi\ket{10} \\  \bra{10}\Psi \rangle \langle\Psi\ket{01} & \bra{10}\Psi \rangle \langle\Psi\ket{10} \end{pmatrix} \notag \\
   &= \begin{pmatrix} (1-\langle IZ \rangle)/2 & (\langle XX \rangle+\langle YY \rangle)/4 \\  (\langle XX \rangle+\langle YY \rangle)/4 & (1-\langle ZI \rangle)/2 \end{pmatrix} \, 
\end{align}
\normalsize

Once measured, this fermionic density matrix contains the effect of noise, and does not represent a pure state. As an effective one-particle pure state, its eigenvalues should be either zero or one.  To project this onto a pure state, we diagonalize $\rho$, identifying the eigenvector $\ket{\phi}$ corresponding to the largest eigenvalue, and then establish a new fermionic density matrix $\rho_p = \ket{\phi}\bra{\phi}$.  This procedure is not scalable for N-fermion problems, but similarly themed attempt have been explored within enforcing N-representability constraints on measured 2-reduced density matrices \cite{Rubin_2018}.  Thus purified energies can then be evaluated using $\rho_p$. 

%

\subsection{Noise Injection}
\begin{figure}[t]
    \centering
    \includegraphics[width=1.0\columnwidth]{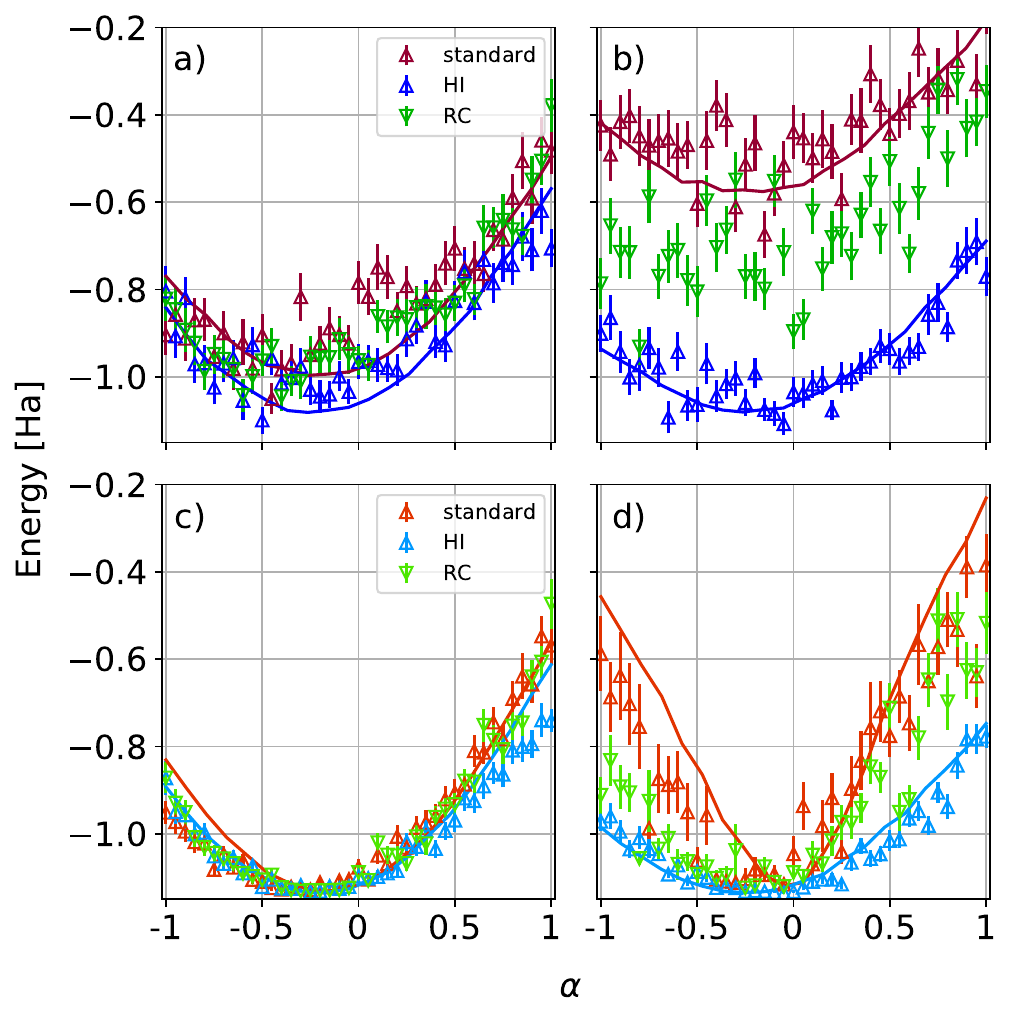}
    \caption{Comparison of HI and RC as mitigation protocols for coherent error showing raw and simulated (straight lines) energies at ideal calibration (a), energies with 0.5 radian under-rotation injection (b), purified energies at ideal calibration (c), and purified energies 0.5 radian under-rotation injection (d).  }  
    \label{fig:two_qubit_coh_error}
\end{figure}

To investigate the noise mitigation properties of these diverse approaches, we intentionally introduce errors onto the two-qubit gates in a variety of manners. While the two-qubit gate in the circuit is the CNOT and its inverse, we inject noise on the native gate of the system, the XX($\pm \pi/2)$ MS gate. As described in section \ref{sec:qscout_exp}, to introduce an under- or over-rotation error, we program that error at the circuit level, resulting in a change in the amplitude of the global beam power delivered. We specifically introduce the rotation error symmetrically, so in the CNOT, we utilize XX($\pi/2 + \epsilon$) and for the CNOT$^\dagger$, XX($-\pi/2-\epsilon$). This is most akin to the types of rotation errors that would occur naturally in the system, which could be due either to a miscalibration of the overall power necessary, or a drift in the overall system detuning also causing systematic over- or under-rotation. In Fig. \ref{fig:two_qubit_coh_error}, we introduce a systematic under-rotation of 0.5 radians (right) in comparison to the nominally fully calibrated approach (left).

\begin{figure}[t]
    \centering
    \includegraphics[width=1.0\columnwidth]{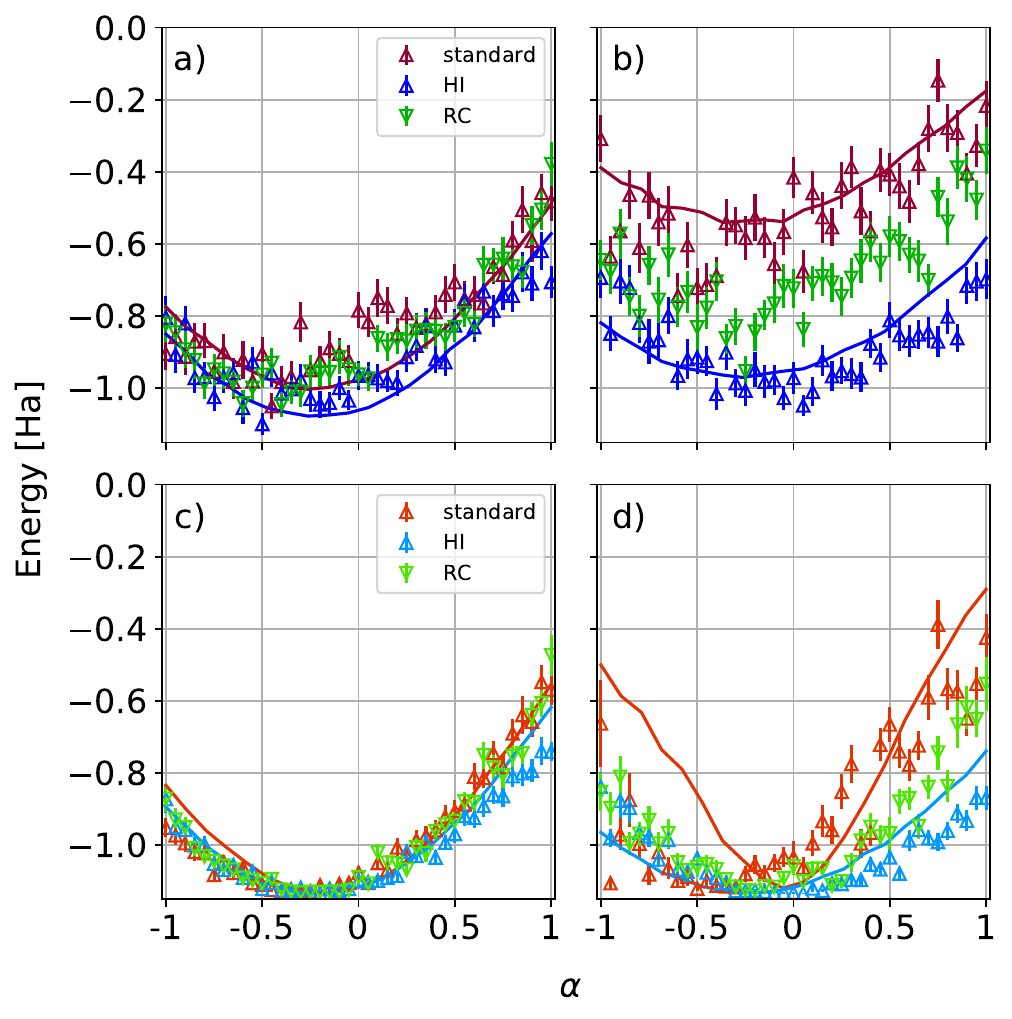}
    \caption{Comparison of HI and RC as mitigation protocols for cooling loop noise, showing raw and simulated energies at ideal calibration with 60 cooling loops (a),energies with only 12 cooling loops (b), purified energies at ideal calibration (c), and purified energies with 12 cooling loops (d).  }   
    \label{fig:two_qubit_heat_error}
\end{figure}

Additionally, we can introduce a broader array of errors through reduction in sideband cooling loops as described in section \ref{sec:qscout_exp}. 
In Fig. \ref{fig:two_qubit_heat_error}, we compare full cooling of 60 loops (left) to a reduction in sideband cooling of 12 loops (right). For full cooling, 60 loops, we see the desired rotation and a residual population in $\ket{01}$ and $\ket{10}$ of $\sim1.5\%$. However, with only 12 cooling loops applied, we see a static under-rotation of $\sim0.3$ radians and population in the $\ket{01}$ and $\ket{10}$ states of $\sim2\%$. Additionally, we expect to see an increase in the stochastic coherent error with a partially cooled ion. In Table \ref{tab:vqe_fids}, we compare the average fidelities of the VQE wave-function (average over all angles) for different error mitigation settings. We find hidden inverses generally outperform randomized compiling for various different injected noise sources.

We performed detailed numerical simulation of the loss landscape with noise injection (solid lines in Fig. \ref{fig:two_qubit_coh_error} and \ref{fig:two_qubit_heat_error}). Our error model consist of ideal single qubit gates, no state prep or measurement errors and all the errors are attributed to the MS gate. Reducing the number of cooling loops affects the effective Rabi frequency and also amplifies the effect of residual entanglement between spin and motion. We model the effective Rabi-frequency change as a combination of static offset to Rabi frequency and a stochastic reduction in Rabi frequency following Debye-Waller effect \cite{Wineland1998NIST}. This can be viewed as a shot to shot coherent error. On the other hand, we model residual entanglement as two-qubit depolarizing noise on the spin states. For the noiseless case, our best fit model has average phonon number of 0.05 (motional mode heating), a static coherent over-rotation of $0.09$ radians (laser intensity miscalibration) and a two-qubit depolarizing probability of 0.02 (residual entanglement between spin and motion) resulting in a 97.5\% MS gate fidelity.  Next, for coherent rotation noise injection, we find a static effective under-rotation of 0.45 from the ideal gate while the two other noise parameters stay the same resulting in a 91\% MS gate fidelity. Finally, for the case where we limit the number of cooling loops to 10-12, we find the best model has average phonon number of 0.5, a static over-rotation of 0.12 radians and a two qubit depolarizing probability of 0.06 resulting in a 89\% MS gate fidelity.



\begin{table*}
    \begin{tabular}{ | l | l | l | l | l | l | l |}
    \hline
    Injected Error & Raw & Pure & HI & HI Pure & RC & RC Pure\\ \hline
    None & 0.923 (33) & 0.996 (3) & 0.950 (24) & 0.997 (2) & 0.943 (22) & 0.997 (2)\\ \hline
    2q 0.25 rad & 0.860 (61) & 0.985 (11) & 0.963 (22) & 0.998 (3) & 0.870 (33) & 0.993 (7)\\ \hline
    2q 0.50 rad & 0.758 (51) & 0.974 (16) & 0.963 (21) & 0.994 (6) & 0.829 (54) & 0.991 (10)\\ \hline
    No Cooling Loops & 0.563 (73) & 0.663 (142) & 0.834 (36) & 0.966 (36) & 0.705 (50) & 0.977 (38)\\ \hline
    12 Cooling Loops & 0.806 (45) & 0.975 (21) & 0.931 (28) & 0.989 (11) & 0.856 (29) & 0.996 (3)\\ \hline
    \end{tabular}
    \caption{Fidelities of VQE wavefunction averaged over all angles.}
    \label{tab:vqe_fids}
\end{table*}



\section{Discussion}
The experimental results and corresponding numerical simulations highlight a few key ideas. First, both hidden inverse gates and randomized compiling are effective methods for coherent noise mitigation. They are markedly different in their application; while randomly compiled circuits add extra twirling gates, the use of HI gates preserves the original circuit depth. They have applicability in different scenarios; for example, an inverse CNOT preceded by a CNOT and a large rotation will not provide the desired noise cancellation. In the work outlined here, we showed that in certain applications with small rotational parameters, such as some instances of VQE, the HI method is very effective near the variational minimum. Secondly, noise amplification allows one to test both the efficacy of a noise mitigation technique and also verify hypotheses about the nature of noise present. Here, noise amplification was used to test the capacity of randomized compiling, HI gates, and density matrix purification to mitigate over and under-rotation errors as well as cooling cycle reduction. These error types can all potentially be attributed to a miscalibrated apparatus (or stale calibrations), which may occur in hardware that has extended up time in between calibration cycles. Third, simulations at the single qubit level were used to fit effective noise model parameters to the experimental data, allowing us to verify hypotheses about the source of the noise as well as the behavior of the noise model under HI circuits. 
In the experimental device, when calibration cycles were explicitly skipped, the model fits predicted that under-rotation errors would be significant source of coherent noise (see Fig.~\ref{Drift}). Explicitly adding under-rotation noise in the experiment further verified the noise model and simulations (and likewise for phase noise), meaning that the noise model adopted here can potentially be used to diagnose potential problems in trapped ion platforms via parameter fitting to a set of data provided by experimentalists. All of the noise mitigation techniques attempted were able to cancel much of the added rotation error as well. Density matrix purification provided a complementary noise mitigation scheme; since it is used in post-processing on classical data provided from measurements, it acts as a ``catch-all'', essentially treating coherent and incoherent noise sources indiscriminately. Because of the method's additional application to stochastic noise and its use in post-processing, it can be effectively used with data provided by either randomly compiled circuits or HI gates. Overall, we find the predictions of our best fit noise model closely align with the experimentally VQE reconstructed loss landscape. Moreover, the parameters of our noise model matches with those estimated in our experimental system. As such, these hidden inverse techniques, in single- and multi-qubit applications, provide a dual purpose, not only error mitigation but also error characterization. 

\section{Conclusion}
Here, we have explored methods of coherent noise mitigation and characterization on a trapped ion quantum processor. We note that low-level noise characterization schemes, consisting of both modeling and noise amplification, combined with noise mitigation techniques, can be used to estimate the sources of noise in trapped ion platforms. Low level characterization techniques, beyond solely measuring gate fidelities, help to make a connection between noise at the gate and quantum control levels and higher level performance. Here, we motivated the use of VQE for quantum chemistry as a key high level application whose performance is correlated with the noise sources studied here. 

\section{Acknowledgments}
Authors thanks Mingyu Kang, Vicente Leyton, and Kenneth Brown for helpful discussions. This material was funded by the U.S. Department of Energy, Office of Science, Office of Advanced Scientific Computing Research Quantum Testbed Program and Quantum Testbed Pathfinder program under ERKJ332. S.M. was supported through US Department of Energy grant DE-SC0019294 awarded to Duke and is funded in part by an NSF QISE-NET fellowship (1747426). Sandia National Laboratories is a multimission laboratory managed and operated by National Technology \& Engineering Solutions of Sandia, LLC, a wholly owned subsidiary of Honeywell International Inc., for the U.S. Department of Energy’s National Nuclear Security Administration under contract DE-NA0003525.  This paper describes objective technical results and analysis. Any subjective views or opinions that might be expressed in the paper do not necessarily represent the views of the U.S. Department of Energy or the United States Government. This work was performed in part at Oak Ridge National Laboratory (ORNL), operated by UT-Battelle for
the U.S. Department of Energy under contract no. DE-
AC05-00OR22725. The United States Government retains and the
publisher, by accepting the article for publication, acknowledges
that the United States Government retains a non-exclusive, paid-up, irrevocable, world-wide license to publish or reproduce the
published form of this manuscript, or allow others to do so, for
United States Government purposes. The Department of Energy will provide public access to these results of federally sponsored research in accordance with the DOE Public Access Plan
(http://energy.gov/downloads/doe-public-access-plan). SAND2022-7158O.

\bibliographystyle{apsrev4-2}
\bibliography{references.bib}
\end{document}